\documentclass[aps,pra,twocolumn,superscriptaddress]{revtex4-2}
\usepackage{amsfonts}
\usepackage{amsmath}
\usepackage{mathrsfs}
\usepackage{amssymb}
\usepackage{graphicx}
\usepackage{bm}
\usepackage{color}
\DeclareMathOperator{\sech}{sech}
\usepackage[colorlinks=true,linkcolor=blue,anchorcolor=blue,citecolor=blue,urlcolor=blue]{hyperref}

\begin{document}

\title{Creating moving gap solitons in spin-orbit-coupled Bose-Einstein condensates}

\author{Jin Su}
\affiliation{International Center of Quantum Artificial Intelligence for Science and Technology (QuArtist) and Department of Physics, Shanghai University, Shanghai 200444, China}
\affiliation{Department of Basic Medicine, Changzhi Medical College, Changzhi 046000, China}

\author{Hao Lyu}
\affiliation{International Center of Quantum Artificial Intelligence for Science and Technology (QuArtist) and Department of Physics, Shanghai University, Shanghai 200444, China}

\author{Yuanyuan Chen}
\affiliation{International Center of Quantum Artificial Intelligence for Science and Technology (QuArtist) and Department of Physics, Shanghai University, Shanghai 200444, China}

\author{Yongping Zhang}
\email{yongping11@t.shu.edu.cn}
\affiliation{International Center of Quantum Artificial Intelligence for Science and Technology (QuArtist) and Department of Physics, Shanghai University, Shanghai 200444, China}

\begin{abstract}
	A simple and efficient method to create gap solitons is proposed in a spin-orbit-coupled spin-1 Bose-Einstein condensate. We find that a free expansion along the spin-orbit coupling dimension can generate two moving gap solitons, which are identified from a generalized massive Thirring model. The dynamics of gap solitons can be controlled by adjusting spin-orbit coupling parameters.
\end{abstract}

\maketitle

\section{Introduction}

It is known that the balance between dispersion and nonlinearity may generate solitons~\cite{Kartashov2011,Konotop2016}. Engineering dispersion and nonlinearity becomes a  merited means to control over the existence of solitons and their dynamics. There is a particular kind of engineered dispersion, which is in the form of an avoided energy level crossing with an energy gap that forbids propagation of linear waves. The balance between such dispersion and nonlinearity places solitons inside the energy gap. Due to their specific location, they are named as gap solitons~\cite{Chen1987,Mills1987}. Generalized massive Thirring models, possessing this dispersion as well as nonlinearity, become an ideal platform to theoretically stimulate gap solitons~\cite{Chang1975,Lee1975,
Christodoulides1989,Aceves1989,Champneys1998,Rossi1998,Mihalache1998}. In experiments, periodic potentials can carry out energy gap opening in dispersion. The observation of gap solitons has been implemented experimentally in various nonlinear periodic optical systems, including fiber Bragg gratings~\cite{Eggleton1996}, waveguide arrays~\cite{Mandelik2004}, optically induced photonic lattices~\cite{Neshev2004,Peleg2004,Lou2007}, temporal lattices~\cite{Bersch2012}, and microcavities with periodic modulations~\cite{Tanese2013}. 

In atomic Bose-Einstein condensates (BECs), optical lattices and spin-orbit coupling are responsible for dispersion engineering~\cite{Eiermann2003,Khamehchi2017}. Energy gaps can be opened up around Brillouin zone edges by optical lattices. Gap solitons are predicted theoretically~\cite{Ostrovskaya2003,Efremidis2003} and observed experimentally~\cite{Eiermann2004} inside these gaps. The study on matter-wave gap solitons in optical lattices soon represents an active research field~\cite{Kartashov2011,Konotop2016,Ahufinger2005,Lee2003,Zhang2009,Zhang2009PRA}. Spin-orbit coupling, which can be implemented experimentally by Raman dressing of atomic hyperfine states~\cite{Lin2011}, can modify dispersion in an interesting way that is different from optical lattices. The spin-orbit-coupled dispersion features many degenerated energy minima and has local  avoided crossings~\cite{Khamehchi2014}. It was predicted long time ago that the local spin-orbit-coupled energy gap can mimic generalized massive Thirring model and can support the existence of gap solitonlike solutions~\cite{Lenz1993,Lenz1994,Zobay1999,Merkl2010,Li2017}. However, until now their experimental observations are not yet achieved. 

A possible experimental challenge may lay in preparation. In order to decorate BECs with spin-orbit coupling, Raman lasers are always ramped up adiabatically in experiments~\cite{Lin2011,Khamehchi2014}. Such loading procedure prepares initial spin-orbit-coupled BECs that are far away from the local energy gaps. It is a challenge for experiments to push initial states into the local energy gaps with high precision. A similar need to transport initial BECs into energy gaps also exists for optical lattices experiments of gap solitons. Precise controllability of optical lattices makes the transport possible. In the experiment, it has been realized by manipulating optical lattices in the way that first accelerating  and then keeping optical lattices to move at a constant velocity~\cite{Eiermann2004}. In spin-orbit-coupled experiments, it is hard to manipulate Raman lasers. This is because that two pseudo-spin states are coupled resonantly by Raman lasers via a two-photon transition. The adjusting of Raman laser frequency gives rise to an extra time-dependent detuning for the resonant transition. The detuning will destroy effects of spin-orbit coupling.

In the present paper, we propose a simple and efficient approach to create gap solitons in spin-orbit-coupled BECs. This method does not need external operations for pushing initial states.  We find that a free expansion triggered by suddenly switching off the harmonic trap along the spin-orbit coupling direction can transfer atoms into the regime of spin-orbit-coupled energy gaps. Atoms inside the regime reach the balance between the dispersion and nonlinearity to form gap solitons.

Our study is in times of rapid advance in the field of spin-orbit-coupled BECs~\cite{Spielman2013,Goldman,Zhai2015,Zhang2016}. Various novel phenomena have been revealed theoretically and experimentally in these systems. The study of solitons in spin-orbit-coupled BECs is attracting much attention. Gap solitons in the local energy gaps can exist for both repulsive and attractive interactions~\cite{Zobay1999, Merkl2010}.  Besides gap solitons, spin-orbit coupling can also support another kind of bright solitons~\cite{Achilleos2013,Xu2013}. These solitons are only produced by attractive interactions and exist in the semi-infinite gap of the spin-orbit-coupled dispersion. Various features of these bright solitons have been addressed theoretically~\cite{Wen2016,Chiquillo2018,Abdullaev2018,Wan2019,Kartashov2020,Zhao2020,Sun2020,Wang2020}. It is interesting to mention that spin-orbit coupling can stabilize bright solitons in high dimensions~\cite{Sakaguchi2014,Salasnich2014,Zhang2015,Mardonov2015,Sakaguchi2016,Xu2015,Li2016,Kartashov2017}, which are always unstable without spin-orbit coupling. On the other hand, optical lattices can open global energy gaps in spin-orbit-coupled dispersion~\cite{Hamner2015}. Therefore, spin-orbit-coupled lattices can accommodate gap solitons and the symmetries of spin-orbit coupling bring them new characters~\cite{Kartashov2013,Zhang2013,Sakaguchi2018}. 

We present a straightforward expansion method to create gap solitons in spin-orbit-coupled spin-1 BECs. The spin-1 spin-orbit coupling has been experimentally implemented in a $^{87}$Rb BEC~\cite{Campbell2016}.  In the spin-orbit-coupled dispersion, there are two independent local avoided crossings sitting at opposite sides in momentum space. Free expansion can push some atoms into both avoided crossings.  Two moving gap solitons can be created and move in opposite direction.  The spin-orbit-coupled spin-1 BECs provide versatile and tunable systems to control over dynamics of gap solitons. We adjust spin-orbit-coupled dispersion to slow down one of them and to make one disappear. 

 The scheme of the paper is organized as follows. In Sec.~\ref{Idea}, we describe the idea of existence of gap solitons in a spin-orbit-coupled spin-1 BEC. We show the spin-orbit-coupled dispersion has two avoided crossings. An effective model including two bands is built up to analyze dynamics around the avoided crossings. From the effective model, we can get analytical gap soliton solutions. In Sec.~\ref{expansion}, we demonstrate that free expansion can dynamically generate two moving bright solitons. The solitons are identified to be gap solitons by fitting soliton profiles with analytical gap soliton solutions.  In Sec.~\ref{Manipulation}, we describe that dynamics of moving gap solitons can be manipulated, and the conclusion follows in Sec.~\ref{Conclusion}.

\section{ Gap solitons in a spin-orbit-coupled spin-1 BEC}
\label{Idea}

Three hyperfine states of $^{87}$Rb can be coupled together by three Raman lasers~\cite{Campbell2016}. Along the Raman lasers propagation direction, there are momentum exchanges between lasers and atoms. The proper manipulation of the momentum exchange generates spin-orbit coupling~\cite{Zhang2016}. Such Raman coupling leads to single-particle spin-orbit-coupled Hamiltonian $H_\text{soc} $ as~\cite{Campbell2016}
\begin{equation}
H_\text{soc} = \frac{1}{2}\left( -i\frac{\partial }{\partial x} +4F_z   \right)^2+\sqrt{2} \Omega F_x +\delta F_z +\epsilon F_z^2,
\end{equation}
with $(F_x,F_y,F_z)$ being spin-1 Pauli matrices.  The Rabi frequency $\Omega$ is proportional to the intensity of Raman lasers. The detuning $\delta$ can be adjusted by changing the frequency of Raman lasers. $\epsilon$ represents the quadratic Zeeman shift and its value depends on the magnitude of the bias magnetic field. The spin-orbit coupling  is denoted by the momentum displacement along the $x$ direction (where Raman lasers propagate) for three hyperfine states. 

\begin{figure}[t]
	\includegraphics[ width=0.5\textwidth]{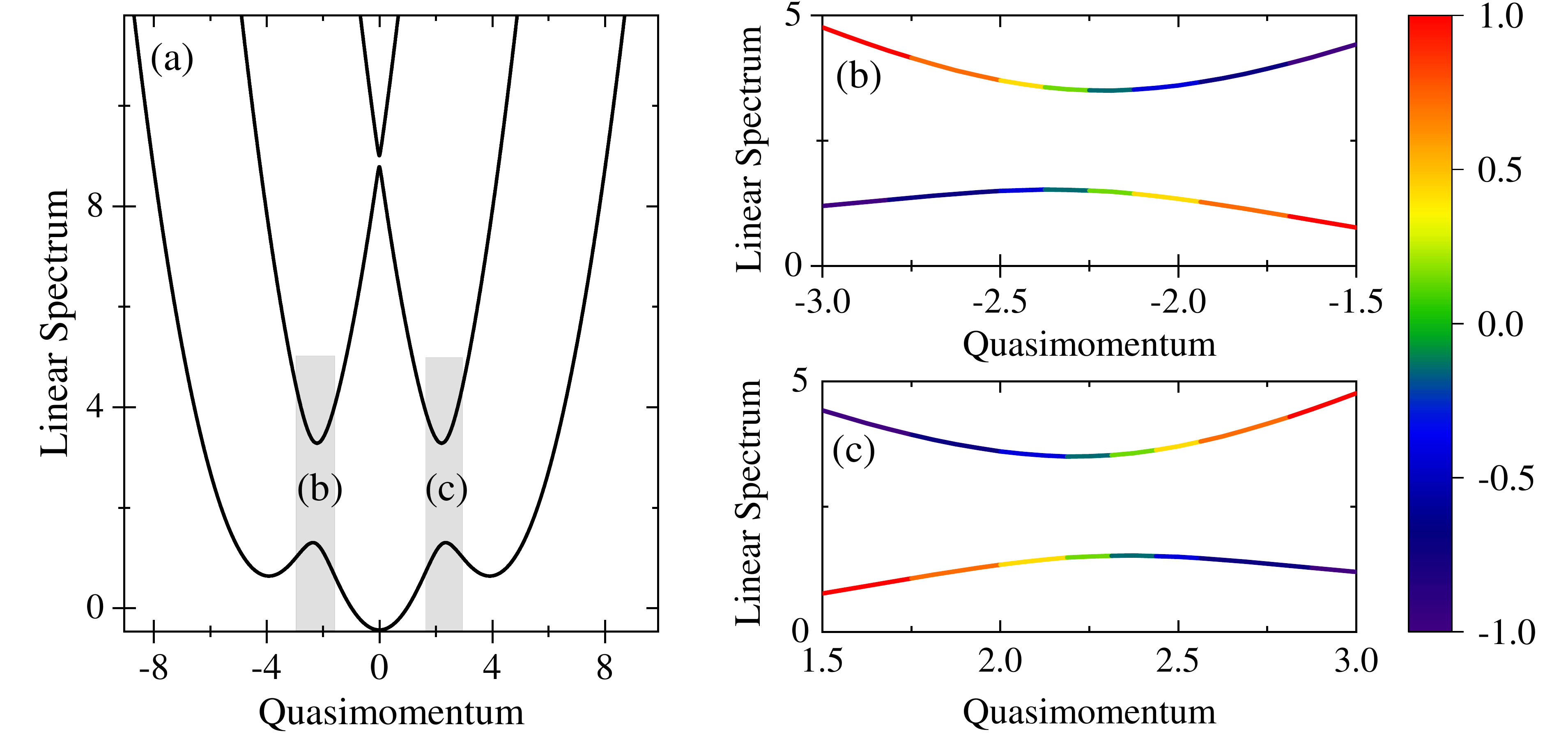}
	\caption{The dispersion relation of the spin-orbit-coupled Hamiltonian $H_\text{soc} $ in the quasimomentum space. The parameter are $\Omega=1$, $\epsilon=1$, and $\delta=0$. (a) The full three bands. Around the smallest separation between the two lower bands, the dispersion appears as two avoided energy crossings, which are shown by the shadow areas.  (b) and (c) are zoom-in of the shadow areas labeled in (a) respectively. Color scales represent the polarization defined as $(|\psi_2|^2-|\psi_1|^2)/(|\psi_1|^2+|\psi_2|^2)$ in (b) and $(|\psi_2|^2-|\psi_3|^2)/(|\psi_2|^2+|\psi_3|^2)$ in (c).    }
	\label{figdispersion}
\end{figure}

\begin{figure*}[t]
	\includegraphics[ width=1\textwidth]{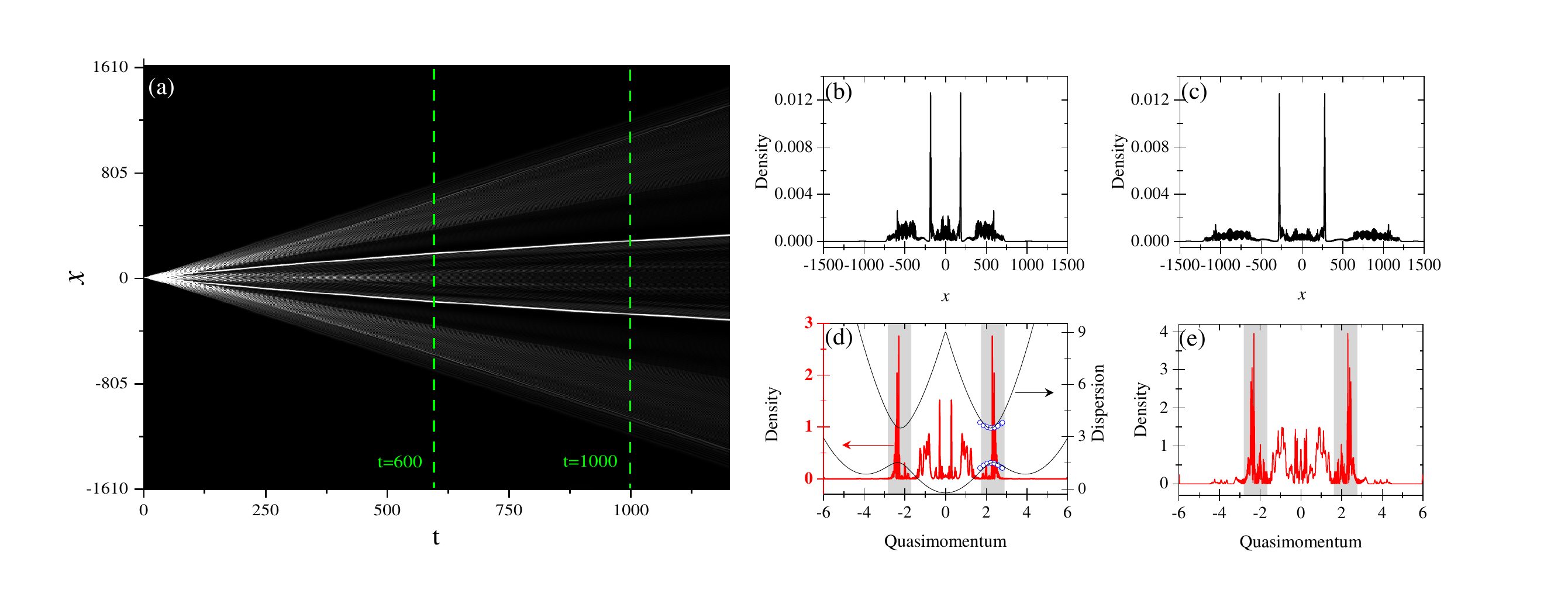}
	\caption{Symmetric expansion triggered by suddenly switching off the longitudinal harmonic trap. The initial state is the ground state with the trap parameter $f_x=0.6$. (a) The evolution of  total density $n(x)=|\psi_1|^2+|\psi_2|^2+|\psi_3|^2$ as a function of $t$. Two sharp density peaks are formed and propagate oppositely with a very long lifetime. Snapshots taken at $t=600$ and $t=1000$ are shown in (b) and (c) respectively. In (d) and (e) the normalized quasimomentum-space densities (red lines) defined in Eq.~(\ref{momentum})  at $t=600$ and $1000$ respectively are shown. Black lines are the  dispersion relation of spin-orbit-coupled Hamiltonian $H_\text{soc}$. Shadowed areas indicate avoided energy crossings featuring negative effective mass in the lowest band.  In (d), circles are fitting to the right avoided crossing. The fitting is $2.5\pm \sqrt{\bar{\gamma}^2(k-2.25)^2+\bar{\Omega}^2} $ with $\bar{\gamma}=1.517$ and $\bar{\Omega}=1$.  The other parameters are $\Omega=1$, $\epsilon=1$, $\delta=0$ and $g_0=10$.}
	\label{figone}
\end{figure*}

Dynamics of a spin-1 BEC with the Raman-induced spin-orbit coupling is described by the Gross-Pitaevskii equation (GPE),
\begin{equation}
\label{GP}
i\frac{\partial \psi }{\partial t}=\left[ H_\text{soc} +V(x)+g_0\left(  |\psi_1|^2+ |\psi_2|^2+ |\psi_3|^2   \right)  \right] \psi.
\end{equation}
The wave functions $  \psi=(\psi_1,\psi_2,\psi_3)^T $ describe occupations of three hyperfine states. The GPE is quasi-one-dimensional, which can be achieved by loading the BEC into an one-dimensional optical dipole waveguide so that motion along the transverse direction $(y,z)$ are frozen.  
For the convenience of numerical calculations, the above GPE is dimensionless. The units of length, energy and time are chosen as $2/k_\text{Ram}$, $\hbar^2k_\text{Ram}^2/4m$ and $4m/(\hbar k_\text{Ram}^2) $, respectively. Here  $m$ is the atom mass and $ k_\text{Ram}=2\pi/\lambda_\text{Ram}$ is the wavevector with $\lambda_\text{Ram}=790$nm  the wavelength of Raman lasers~\cite{Campbell2016}. 
The harmonic trap is $V(x)= f_x^2 x^2/2$ with dimensionless $f_x$ relating to the trap frequency $\omega_x$,  $f_x=2\sqrt{2}m\omega_x/(\hbar k_\text{Ram}^2 )$. The interaction coefficient in Eq.~(\ref{GP}) is $g_0=4Na_0m \omega_r/( \hbar k_\text{Ram} )$  with $N$ being the atom number, $a_0$ being the spin-independent scattering length and $\omega_r$ being the trap frequency along the transverse direction. For a $^{87}$Rb BEC,   $a_0=100a_B$ with $a_B$  the Bohr radius. We neglect the spin-dependent scattering length $a_2$ since it is very small in experiments~\cite{Campbell2016}.
The wave functions satisfy normalization condition, $\int dx ( |\psi_1|^2+ |\psi_2|^2+|\psi_3|^2 )=1$.

The dispersion relation of  the spin-orbit-coupled Hamiltonian $H_\text{soc}$ has interesting features. A typical dispersion is shown in Fig.~\ref{figdispersion}.  In quasimomentum $k$ space, there are three parabola locating at $k=\pm 4$ and $0$ respectively. Local gaps are opened by the Rabi frequency at crossings between parabola. At last, the dispersion relation has three bands. The lowest band possesses three energy local minima. The bias between them can be tuned by changing  the detuning and quadratic term. Around the smallest separations between the two lower bands, the dispersion appears as two local avoided energy crossings.  In the regimes of avoided crossings [shown by the shadow areas in Fig.~\ref{figdispersion}(a)], the lowest band possesses negative effective mass. The left avoided crossing is dominated by the components $\psi_1$ and $\psi_2$, and the involvement of the component $\psi_3$ is very small. This is can be seen from the polarization
$(|\psi_2|^2-|\psi_1|^2)/(|\psi_1|^2+|\psi_2|^2)$ demonstrated by color scales in Fig.~\ref{figdispersion}(b) for the left avoided crossing. While  the right avoided crossing is mainly occupied by the components $\psi_2$ and $\psi_3$, and the population of the component $\psi_1$ can be neglected. The polarization for the right avoided crossing is defined as, $(|\psi_2|^2-|\psi_3|^2)/(|\psi_2|^2+|\psi_3|^2)$, which is demonstrated in Fig.~\ref{figdispersion}(c).

Since the local avoided crossings are far from the higher band and are dominated  by two components, if physics is only relevant to the avoided crossings, they can be effectively described by the Hamiltonian,
\begin{equation}
H_\text{eff}=\bar{\gamma}\sigma_z (-i \frac{\partial }{\partial x}-k_\text{cen}) +\bar{\Omega} \sigma_x,
\end{equation}
with eigenstates $\phi=(\phi_a, \phi_b )^T$ being two components.  $\sigma_z$ and $\sigma_x$ are spin-1/2 Pauli matrices. The dispersion relation of $H_\text{eff}$ is  $\pm \sqrt{ \bar{\gamma}^2(k-k_\text{cen})^2+\bar{\Omega}^2}$, having the form of avoided crossing in the quasimomentum space. $\bar{\Omega}$ describes the gap size of the avoided crossing, $\bar{\gamma}$ depicts its steepness, and $k_\text{cen}$ is the location where the avoided crossing is centered. 

The avoided crossings, existing in the spin-orbit-coupled Hamiltonian $H_\text{soc}$, provide a versatile bed for the investigation of gap solitons~\cite{Lenz1993,Lenz1994,Zobay1999}. Assuming atoms are confined around an avoided crossing, their dynamics may be described by the effective model as,
\begin{equation}
\label{Effective}
i\frac{\partial\phi}{\partial t}  = H_\text{eff} \phi + g_0 (| \phi_a|^2+|\phi_b|^2)\phi.
\end{equation}
In above equation,  interactions between atoms are considered by the last term. The effective  model, in literature,  is known as a generalized massive Thirring model~\cite{Chang1975,Lee1975,
Christodoulides1989,Aceves1989,Champneys1998,Rossi1998,Mihalache1998}.  The massive Thirring model, sharing the same $H_\text{eff}$ but having specific nonlinearity, is completely integrable~\cite{Thirring1958}. The effective model in Eq.~(\ref{Effective}) includes a generalized nonlinearity.  It is known that the outstanding feature of the model is to admit analytical moving gap soliton solutions~\cite{Christodoulides1989,Aceves1989,Lenz1993,Lenz1994,Zobay1999,Sterke1990}. They are,




\begin{equation}
\label{soliton}
\begin{pmatrix}
\phi_a\\ \phi_b
\end{pmatrix}=Ae^{-ixk_\text{cen}}
\begin{pmatrix}
\ \  \frac{1}{\Delta} \sech\left( \theta-i\frac{\Gamma}{2} \right) \\
-\Delta  \sech\left( \theta+i\frac{\Gamma}{2} \right)
\end{pmatrix},
\end{equation}
with
\begin{equation}
A=\sqrt{\frac{\bar{\Omega}(1-v^2)}{2g}} \left(  -\frac{e^{2\theta}+e^{-i\Gamma}  }{e^{2\theta}+e^{i\Gamma}}  \right)^ve^{i\sigma }\sin\Gamma,\notag 
\end{equation}
\begin{equation}
\theta=\frac{\bar{\Omega}  (-x/ \bar{\gamma}+vt)\sin\Gamma}{\sqrt{1-v^2}},  \ \ \sigma=\frac{\bar{\Omega} (-vx/ \bar{\gamma}+t) \cos\Gamma}{\sqrt{1-v^2}}, \notag
\end{equation}
and
\begin{equation}
\Delta=\left( \frac{1-v}{1+v}  \right)^{\frac{1}{4}}.\notag 
\end{equation}
Two parameters $v$ ($-1<v<1$) and $\Gamma$ ($0<\Gamma <\pi$) determine the velocity and amplitude of gap solitons.

These moving gap solitons are supported by the avoided crossing dispersion. For the spin-orbit-coupled spin-1 system, two avoided crossings sit at opposite sides with $k_\text{cen}\ne 0$ and are far from $k=0$ [as shown in Fig.~\ref{figdispersion}(a)]. In experiments,  to artificially introduce spin-orbit coupling into the BEC,  Raman lasers should be ramped up adiabatically, in this way, the BEC is always prepared in spin-orbit-coupled ground state which  locates at the global minimum of lowest band in the dispersion relation of $H_\text{soc}$ (i.e., $k=0$). In order to create moving gap solitons, the transport of initially prepared BEC from $k=0$ to  $k_\text{cen}$ is prerequisite. However, the precise manipulation of the transport in experiments may be a challenge. In the following, we reveal that a free expansion could be an experimentally accessible way to create moving gap solitons.

\begin{figure}[t]
	\includegraphics[ width=0.4\textwidth]{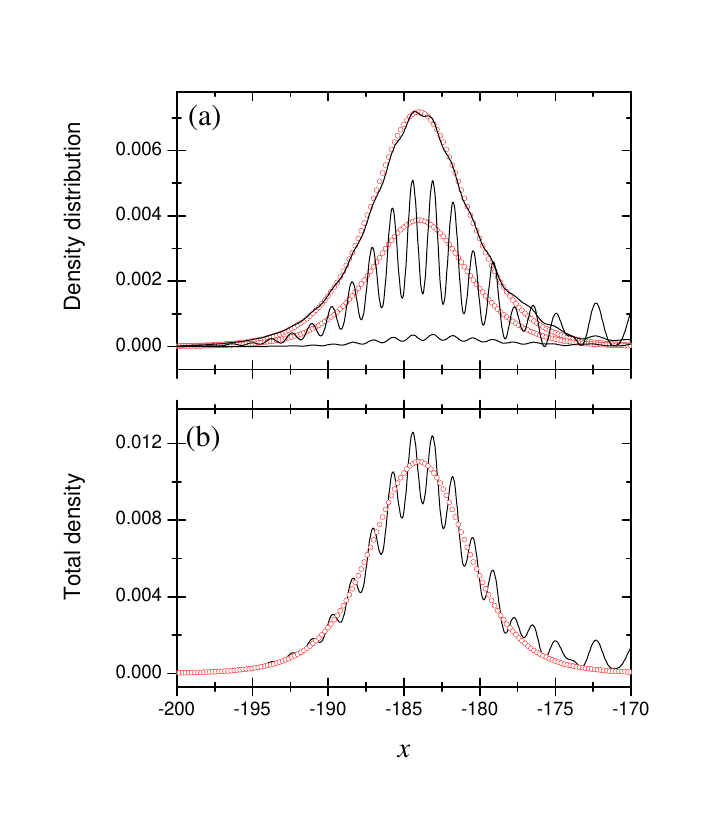}
	\caption{Matching of density distributions between the negative-direction moving peak taken at $t=600$  in Fig.~\ref{figone} and an analytical soliton solution from Eq.~(\ref{soliton}). All parameters are same as in Fig.~\ref{figone}.  In (a), each component density is presented.  Black solid lines from top to bottom are  $|\psi_3|^2$, $|\psi_2|^2$ and $|\psi_1|^2$.  Red open-circle lines are gap soliton solution.  Upper and lower ones correspond to $|\phi_b|^2$ and $|\phi_a|^2$ respectively. In (b), total density distribution $|\psi_1|^2+|\psi_2|^2+|\psi_3|^2$ (black solid line) and  $|\phi_a|^2+|\phi_b|^2$ (red open-circle line) are shown.  The parameters of the avoided crossing are $\bar{\gamma} =1.517$ and $\bar{\Omega}=1$. The parameters of analytical gap solitons are $v=-0.306$ and $\Gamma=0.342$. }
	\label{figtwo}
\end{figure}

\begin{figure}[t]
	\includegraphics[ width=0.45\textwidth]{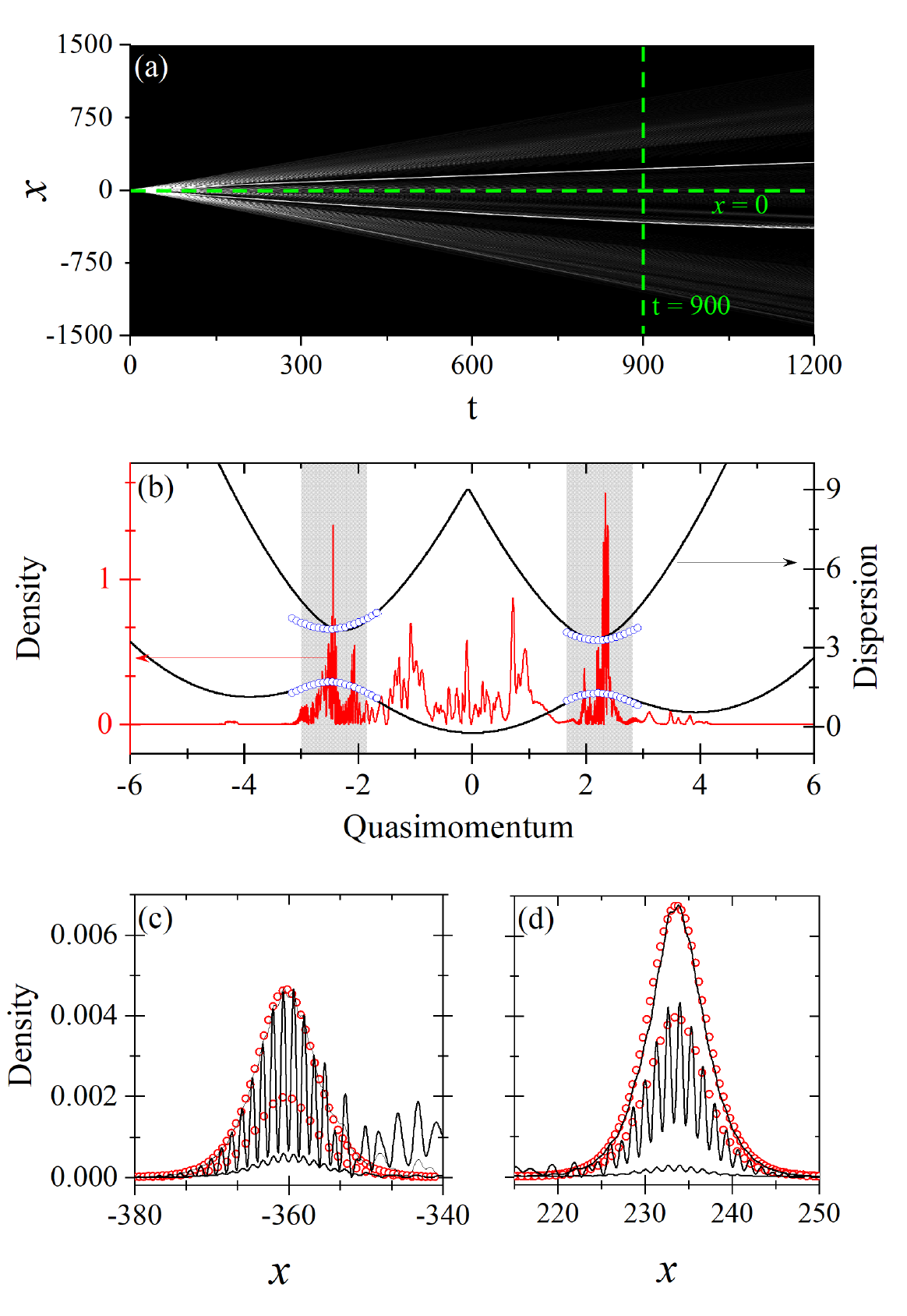}
	\caption{ Asymmetric expansion with a nonzero detuning $\delta=0.3$. The detuning gives rise to the asymmetry of two avoided crossing with respect to $k=0$.  (a) Total density evolution.  (b) The normalized quasimomentum-space denisty (red line) at $t=900$ and the dispersion relation of $H_\text{soc}$ (black lines).  The shadow areas indicate avoided crossings, the fittings of which are represented by circles. We use the fitting as $2.28\pm \sqrt{\bar{\gamma}^2(k-2.2)^2+\bar{\Omega}^2}$ for the right avoided crossing and $2.7\pm \sqrt{\bar{\gamma}^2(k+2.5)^2+\bar{\Omega}^2}$ for the left one with  $\bar{\gamma} =1.517$ and $\bar{\Omega}=1$. (c,d) Matching of  expansion result at $t=900$  (black solid lines) and analytical gap soliton solutions in Eq.~(\ref{soliton}) (red open-circle lines). (c) Soliton moves in negative direction with $v=-0.4$ and $\Gamma=0.27$. Black solid lines from top to bottom correspond to $|\psi_3|^2$, $|\psi_2|^2$ and $|\psi_1|^2$. Red open-circle lines from upper to lower are $|\phi_b|^2$ and $|\phi_a|^2$.  (d) Soliton moves in positive direction with $v=0.3$ and $\Gamma=0.335$.  Black solid lines from top to bottom correspond to $|\psi_1|^2$, $|\psi_2|^2$ and $|\psi_3|^2$. Red open-circle lines from upper to lower are $|\phi_a|^2$ and $|\phi_b|^2$. The other parameters are $\Omega=1$, $\epsilon=1$ and $g_0=10$.}
	\label{figthree}
\end{figure}

\section{ Creating moving gap solitons by expansion}
\label{expansion}

We start from a spin-orbit-coupled ground state obtained by numerically solving Eq.~(\ref{GP}) using the imaginary time evolution. The harmonic trap $V(x)$ is then suddenly switched off.  The quench allows  the BEC to expand along the longitudinal direction $x$ in the presence of spin-orbit coupling.  The interactions between atoms are converted into driving forces to move atoms. If quench-induced excitations are weak comparing with the band separation between the two lower bands in the dispersion relation, the expansion mainly involves the lowest band~\cite{Khamehchi2017}.

The expansion of total density, $n(x)=|\psi_1(x)|^2+|\psi_2(x)|^2+|\psi_3(x)|^2$, is presented in Fig.~\ref{figone}(a). In coordinate space, the expansion of the total density is symmetric with respect to $x=0$. Around $t=150$, two sharp symmetric density peaks appear. They move oppositely at a constant velocity for a very long time and keep their height unchanged. Fig.~\ref{figone}(b) and \ref{figone}(c) demonstrate two snapshots of the evolution, which are taken at $t=600$ and $t=1000$ respectively. Beside the density peaks, there are other excited modes. As the other modes propagate, the density peaks become more obvious.  The chosen parameters result in a symmetric  dispersion relation with respect to $k=0$ [shown in Fig.~\ref{figone}(d)]. Initial state is centered around $k=0$. During expansion, quasimomentum is symmetrically extended in opposite directions. As quasimomentum increases, some atoms enter into negative effective mass regimes of lowest band [indicated by shadowed areas in Fig.~\ref{figone}(d) and \ref{figone}(e)]. When atom number inside these regimes is apparently larger than other occupation of quasimomentum, the density peaks are emergent in coordinate space. Fig.~\ref{figone}(d) and \ref{figone}(e) show two snapshots of normalized quasimomentum-space density distributions. The normalized quasimomentum-space density is defined as,
\begin{equation}
\label{momentum}
\frac{|\psi_1(k)|^2+|\psi_2(k)|^2+|\psi_3(k)|^2}{\int dk \left[ |\psi_1(k)|^2+|\psi_2(k)|^2+|\psi_3(k)|^2   \right]},
\end{equation}
with $|\psi_i(k)|^2$ being the density of $i$th component in quasimomentum space. From these plots, it is clear that the occupation of negative effective mass regimes completely dominates. The appearance of the sharp density peaks in coordinate space always accompanies the dominant occupation of negative effective mass regimes. It looks like that some atoms are ``trapped"  by negative effective mass regimes. Such trapped states can sustain for a long time, this requires a proper range of interaction coefficient: if the coefficient is small, there will be not enough driving resource for expansion to push atoms into negative effective mass regimes; if  it is large, atoms in negative effective mass regimes shall be continuously moved out by further expansion.

\begin{figure}[t]
	\includegraphics[ width=0.4\textwidth]{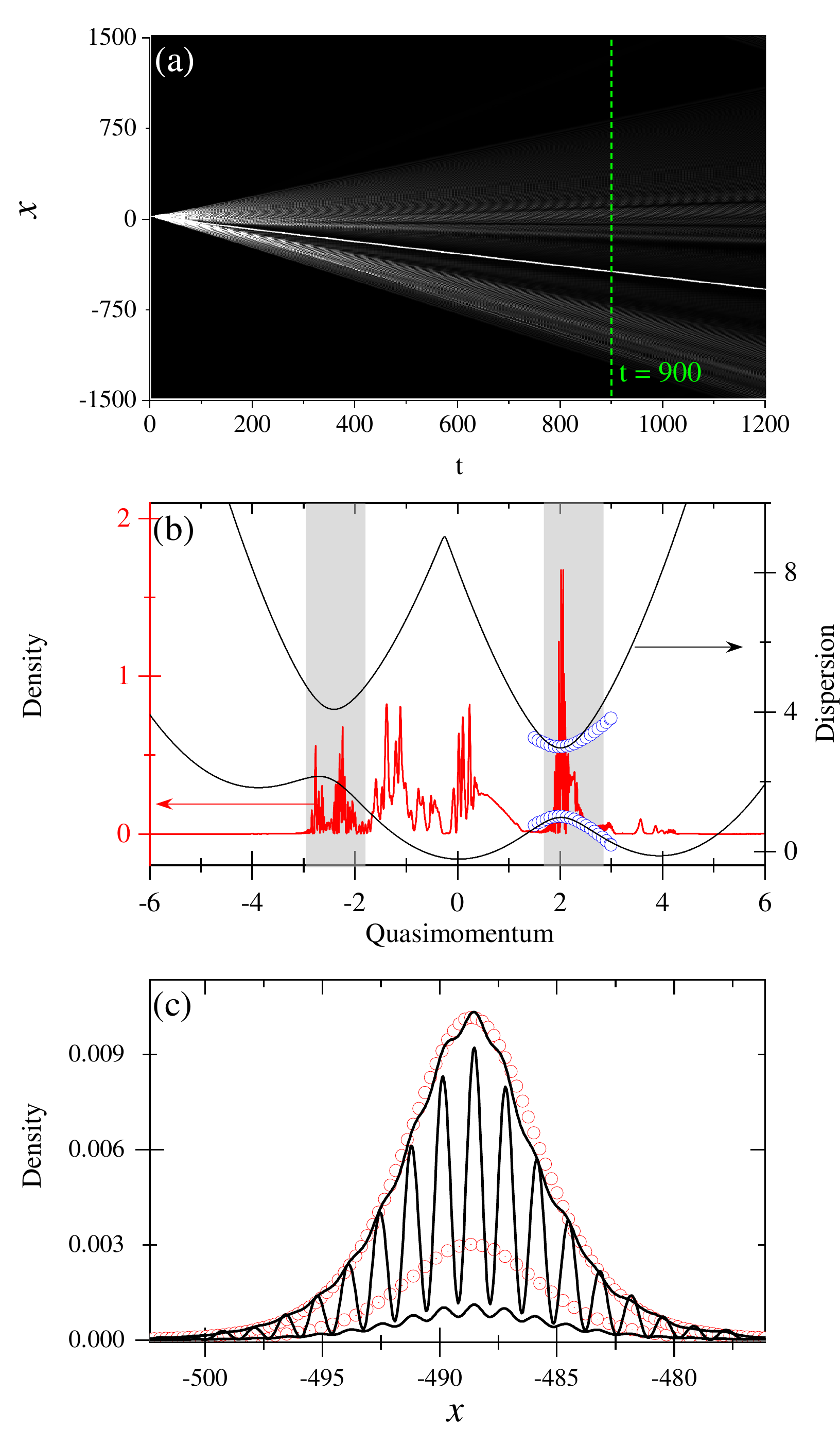}
	\caption{Asymmetric expansion with $\delta=1$. All panels show same quantities as in Fig.~\ref{figthree}. But here $g_0=8$.  The fitting in (b) is $2 \pm \sqrt{ \bar{\gamma}^2(k-2)^2+\bar{\Omega}^2}$ with $\bar{\gamma} =1.517$ and  $\bar{\Omega}=1$. In (c),  parameters of the gap soliton solution are $v=0.54$  and $\Gamma=0.36$. Black solid lines from top to bottom correspond to $|\psi_3|^2$, $|\psi_2|^2$ and $|\psi_1|^2$. Red open-circle lines from upper to lower are $|\phi_b|^2$ and $|\phi_a|^2$. }
	\label{figfour}
\end{figure}

The longevous sharp density peaks in coordinate space in Fig.~\ref{figone}(a) are identified as moving gap solitons. Nearby the negative effective mass regimes, the two lower bands are close to each other and form avoided crossings. The effective model in Eq.~(\ref{Effective}) provides a simple way to describe atoms inside the regimes. To precisely mimic the avoided crossings in the dispersion of $H_\text{soc}$, we first fix the parameters $\bar{\gamma}, \bar{\Omega}$ and $k_\text{cen}$ in $H_\text{eff}$ by fitting the dispersion of $H_\text{eff}$ to the avoided crossings. The fitting of the right avoided crossing as an example is demonstrated in Fig.~\ref{figone}(d). We shall reveal that the sharp density peaks  are moving gap solitons in Eq.~(\ref{soliton}).  For this purpose, we extract information of the velocity and amplitude of the density peaks. Then we apply them to the analytical gap soliton solutions to fix the free parameters $v$ and $\Gamma$ in Eq.~(\ref{soliton}).  In Fig.~\ref{figtwo}, we show the negative-direction moving density peak together with the profiles of gap solitons. The density peak is taken at $t=600$ in Fig.~\ref{figone}(a).  On account of the mismatch of time origin for the density peak and gap solitons, we choose a $t$ that is different from $600$ for gap solitons. With this $t$, the center of gap soliton coincides with the density peak. It is shown that profiles of gap solitons match with the density peak very well.  Therefore, the density peak is a gap soliton and originates from the occupation of the right  avoided crossing. As the right avoided crossing mainly comes from the coupling between $\psi_2$ and $\psi_3$,  the occupation of the first component $\psi_1$ [the bottom solid line in Fig.~\ref{figtwo}(a)] is very small.  We notice that spatial modulation exists in the density peak, which is more evident in the second component $\psi_2$ [the middle solid line in Fig.~\ref{figtwo}(a)] . This is because beside the dominant occupation of avoided crossing regimes the expansion also produces some modes around $k=0$ [see Fig.~\ref{figone}(d) and \ref{figone}(e)]. These modes are primarily composed of the second component $\psi_2$, and are occupied by just a few atoms. The interference between them and the density peaks contributes to the modulation on the top of the density peaks. The total density $n(x)$ also shows modulation, and gap soliton solution can fit it very well, which is demonstrated in  Fig.~\ref{figtwo}(b).

We conclude that the density peaks are moving gap solitons, which can be created through a free expansion. An initial spin-orbit-coupled BEC can be pushed to occupy the regime of avoided crossings in dispersion relation, where atoms form gap solitons.

\section{ Manipulating moving gap solitons}
\label{Manipulation}

In this section, we demonstrate that the spin-orbit-coupled spin-1 system is very versatile for controlling dynamics of gap solitons. It can be achieved by adjusting the detuning $\delta$ in $H_\text{soc}$.  In the previous section, two symmetric gap solitons can be created with a zero detuning $\delta=0$.  We reveal that in the presence of the detuning two gap solitons become asymmetric: for a small detuning, two gap solitons have different velocity; for a large detuning one of them disappears. 

Expansion of the total density with $\delta=0.3$ is presented in Fig.~\ref{figthree}(a). The expansion can still generate two moving gap solitons. However, the negative-direction moving one has a larger velocity. In the presence of the detuning, the dispersion relation of  $H_\text{soc}$ becomes asymmetric with respect to $k=0$. The right avoided crossing is energetically lower than the left one, and during the increase of quasimomentum in the expansion more atoms are sent into the right regime as shown in Fig.~\ref{figthree}(b).  In Fig.~\ref{figthree}(c) and \ref{figthree}(d), we match expansion results with analytical gap soliton solutions in Eq.~(\ref{soliton}).  The good match in Fig.~\ref{figthree}(c) and \ref{figthree}(d) verifies that the expansion with a small detuning induces asymmetric gap solitons. Oscillations in $\psi_2$ still exist due to the excited modes around $k=0$.

Expansion of the total density with a larger detuning ($\delta=1$) is shown in Fig.~\ref{figfour}(a). There is only one soliton produced from background and it moves in the negative direction. With this detuning, the right avoided crossing in the dispersion relation of $H_\text{soc}$ becomes more obviously lower than the left one [see Fig.~\ref{figfour}(b)]. The expansion can easily delivers atoms into the right regime.  This causes that there are no enough atoms in the left regime to form gap solitons. Therefore, only atoms in the right  regime can form soliton. The density of soliton in Fig.~\ref{figfour}(c) shows that populations of $\psi_2$ and  $\psi_3$ dominate, which confirms that the soliton comes from the occupation of the right avoided crossing.  The match presented in Fig.~\ref{figfour}(c) proves that the soliton is a gap soliton.  It should be noted that the interaction coefficient $g_0$ is smaller than that in Fig.~\ref{figthree}. We numerically find that to excite only one soliton the range of $g_0$ would be $7.0<g_0<9.3$, which corresponds to the total atom number $2000<N<2660$ if the typical transverse tap $\omega_r=2\pi \times 150 $Hz is considered.

\section{Conclusion}
\label{Conclusion}

We present the straightforward free expansion method to create gap solitons in an experimentally realizable spin-orbit-coupled spin-1 BEC. Via expansion, some atoms can be delivered into the regime of spin-orbit-coupled energy gaps to form gap solitons. With this method, the created gap solitons are movable.  The idea of our study can be applied to a spin-orbit-coupled spin-1/2 BEC, where only one moving gap solitons is expected to be created.

\begin{acknowledgments}
	
 This work is supported by National Natural Science Foundation of China with Grants Nos.~11974235 and 11774219.
	
\end{acknowledgments}

\end{document}